\documentclass[aps,prl,superscriptaddress,showpacs,floatfix,nobibnotes]{revtex4}
\usepackage{epsfig}
\usepackage{epstopdf}

\usepackage{graphicx}
\usepackage{longtable}
\usepackage{CJK}
\usepackage{color}

\usepackage{mathptmx, courier, pifont}
\usepackage[scaled=0.92]{helvet}
\usepackage[T1]{fontenc}
\usepackage{textcomp}

\begin{document}

\title{Anomalous low-energy $E2$-related behavior in triaxial nuclei}

\author{Yu Zhang}\email{dlzhangyu_physics@163.com}
\affiliation{Department of Physics, Liaoning Normal University,
Dalian 116029, P. R. China}

\author{Ying-Wen He}
\affiliation{Department of Physics, Liaoning Normal University,
Dalian 116029, P. R. China}

\author{D. Karlsson}
\affiliation{Department of Physics, KTH Royal Institute of
Technology, Stockholm 10691, Sweden}

\author{Chong Qi}\email{chongq@kth.se}
\affiliation{Department of Physics, KTH Royal Institute of
Technology, Stockholm 10691, Sweden}

\author{Feng Pan}\email{daipan@dlut.edu.cn}
\affiliation{Department of Physics, Liaoning Normal University,
Dalian 116029, P. R. China}\affiliation{Department of Physics and
Astronomy, Louisiana State University, Baton Rouge, LA 70803-4001,
USA}

\author{J. P. Draayer}\email{ draayer@lsu.edu}
\affiliation{Department of Physics and Astronomy, Louisiana State
University, Baton Rouge, LA 70803-4001, USA}

\date{\today}

\begin{abstract}
The anomalous low-energy $E2$-related behavior of a
triaxially-deformed nucleus bas been identified and analyzed based on the SU(3) algebraic 
theory within the framework of the interacting boson model.
The results show striking features that include a
$B(E2;4_1^+\rightarrow2_1^+)/B(2_1^+\rightarrow0_1^+)<1.0$ transition rate ratio and a
$E(4_1^+)/E(2_1^+)>2.0$ excitation energy ratio
that can be tracked back to a finite-$N$ effect,
which in a large-$N$ limit of the theory yields normal
results for a stable $\gamma$-deformation. This description
is shown to be able to explain observed $E2$ anomalous phenomenon in neutron-deficient nuclei
such as $^{172}$Pt and $^{168}$Os,
and in so doing yields a deeper understanding of the physical features of a soft triaxially-deformed nucleus.
\end{abstract}


\maketitle

The emergence of collective features is one of the most
important and striking characteristics of complex nuclear many-body
systems. How nuclear collectivity emerges from collective modes
(shapes and deformations thereof) can be realized theoretically from
both macroscopic and microscopic perspectives. Specifically, various
collective modes can be explained within a Bohr-Mottelson picture of
the dynamics using geometric language~\cite{Bohrbook}, which includes
the notion of a spherical vibrator, that of either an axially-deformed
or a triaxially-deformed rotor of the Davydov and Filippov~\cite{DF} type,
or even the $\gamma$-unstable rotational motion introduced by Wilets and
Jean~\cite{WJ} which stands in sharp contrast with that of a rigid or
$\gamma$-stable characterization of the dynamics.

On the other hand, in contrast with the above, the interacting boson
model (IBM)~\cite{IachelloBook87}, which is an algebraic theory, has also
demonstrated excellent success in elucidating collective nuclear modes.
A big advantage of the IBM is that not only are there three distinguishable
collective limits that can be realized, but even more importantly, one can use
the theory to study the results of the mixing of modes, and as well the
fact that the spatial reach or size of the model space can be controlled by
the number of bosons that are allowed to participate in the dynamics. In
what follows below, we capitalize on this flexibility of the IBM, and in
particular show that the boson-number dependence on the low-energy
$E2$-related behavior of triaxially deformed nuclei can be used to explain the origin
of the anomalous low-energy $E2$-related behavior that has recently been observed
in some heavy neutron-deficient nuclei.

How to understand nuclear collectivity in terms of its microscopic roots
is certainly a highly desirable proposition, but such approaches are usually
plagued by various high levels of complexities that often include in addition
major computational challenges. Nevertheless, it should be noted that much
progress toward such a goal has been made based on so-called {\it ab~initio}
shell-model theories. The seminal work of Elliott~\cite{Elliott1958} opened
the door to a fully microscopic pathway for achieving a truly microscopic
understanding of rotational motion in light nuclei. Furthermore, some major
steps forward in this direction
have been achieved more recently using what has been dubbed by its founders as
a no-core shell model theory, through which the emergent collectivity in light
nuclei and the symmetries that underpin such modes have been addressed starting with
{\it ab~initio} (from first principles) interactions that are parameter free
\cite{Draayer2012,Barrett2013,Dytrych2013,Dytrych2020,McCoy2020}. Also, and
much earlier than the latter, for heavy systems the so-called pseudo-SU(3)
shell model was advanced and shown to be able to successfully describe many
features of low-lying collective phenomena in strongly deformed heavy nuclei
~\cite{Draayer1983}. In addition, the proxy-SU(3) scheme has also been proffered as
a similar methodology~\cite{Bonatsos2017} along with some analytic results
for considering associated collective features. All these developments, from Elliott and
forward, including the SU(3) limit of the IBM - suggest that SU(3) - which is the symmetry group of the 3D-Harmonic Oscillator - plays
an essential role in attempts to gain a deeper understanding of the origin of
collectivity from a more microscopic perspective.

Overall, the nature of collective modes in atomic nuclei is best revealed through the low-lying
spectroscopic features of nuclei and the associated $E2$-related electromagnetic
transitions. However, recent experimental
measurements~\cite{Grahn2016,Saygi2017,Cederwall2018,Goasduff2019,Zhang2021}
suggest a puzzling anomalous phenomenon among some low-lying yrast states
in certain neutron-deficient nuclei including $^{166}$W, $^{168,170}$Os
and $^{172}$Pt, within which the excitation energy ratio $R_{4/2}\equiv
E(4_1^+)/E(2_1^+)>2.0$ shows the collective nature of these states while the results are accompanied by rare and anomalous $B(E2)$ ratio with
$B_{4/2}\equiv B(E2;4_1^+\rightarrow2_1^+)/B(E2;2_1^+\rightarrow0_1^+)<1.0$.
This anomalous behavior seems to persist in the neighboring odd-$A$ nuclei,
for example in $^{169}$Os, with an odd neutron outside of an even-even
core serving as a spectator. What is clear is that this well-documented phenomenon does not seem
to belong to any particular set of the more familiar conventional collective modes, nor has it been addressed in a convincing way in large-scale shell model approaches~\cite{Cederwall2018}
or within self-consistent mean-field analyses~\cite{Goasduff2019}.

The main
purpose of this paper is to tackle this problem within the IBM framework,
and in particular, within its representation in terms of the SU(3) symmetry limit of the theory.
Specifically, in what follows we show how this novel collective feature
emerges naturally within the SU(3) realization of a triaxially-deformed rotor, when
a finiteness effect is added to the theory, which yields $R_{4/2}>2.0$ ratios and
$B_{4/2}<1.0$ values simultaneously, and in so doing this picture provides a
relatively simple explanation for the observed anomaly. We also proffer that it seems
reasonable to suggest that this feature should as well as be
found in other models that incorporate SU(3)-defined basis states
~\cite{Draayer1983,Bonatsos2017,Rowe1985,Georgieva1983,Wu1987,Zuker1995}.

\begin{figure}
\begin{center}
\includegraphics[scale=0.21]{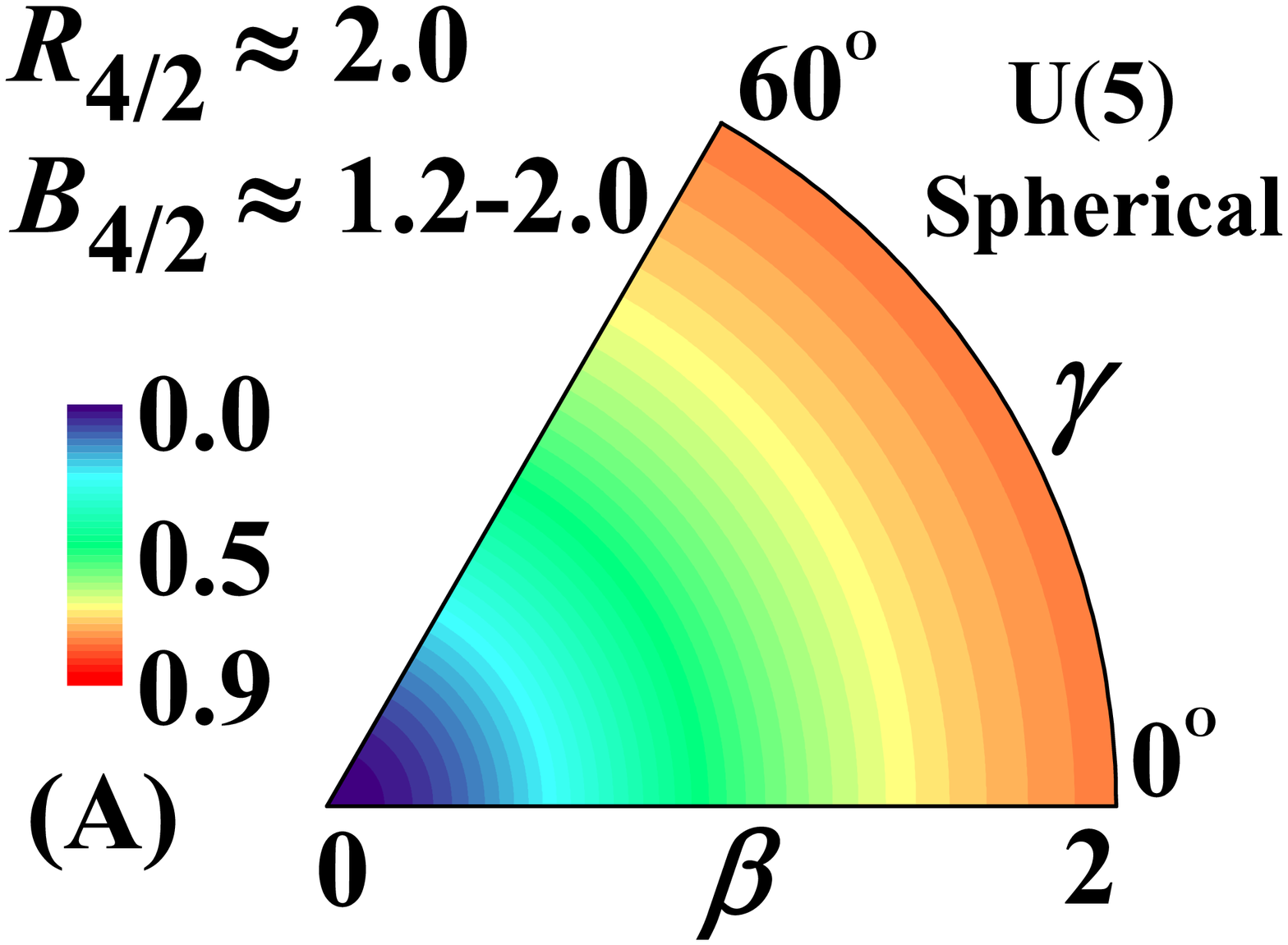}
\includegraphics[scale=0.21]{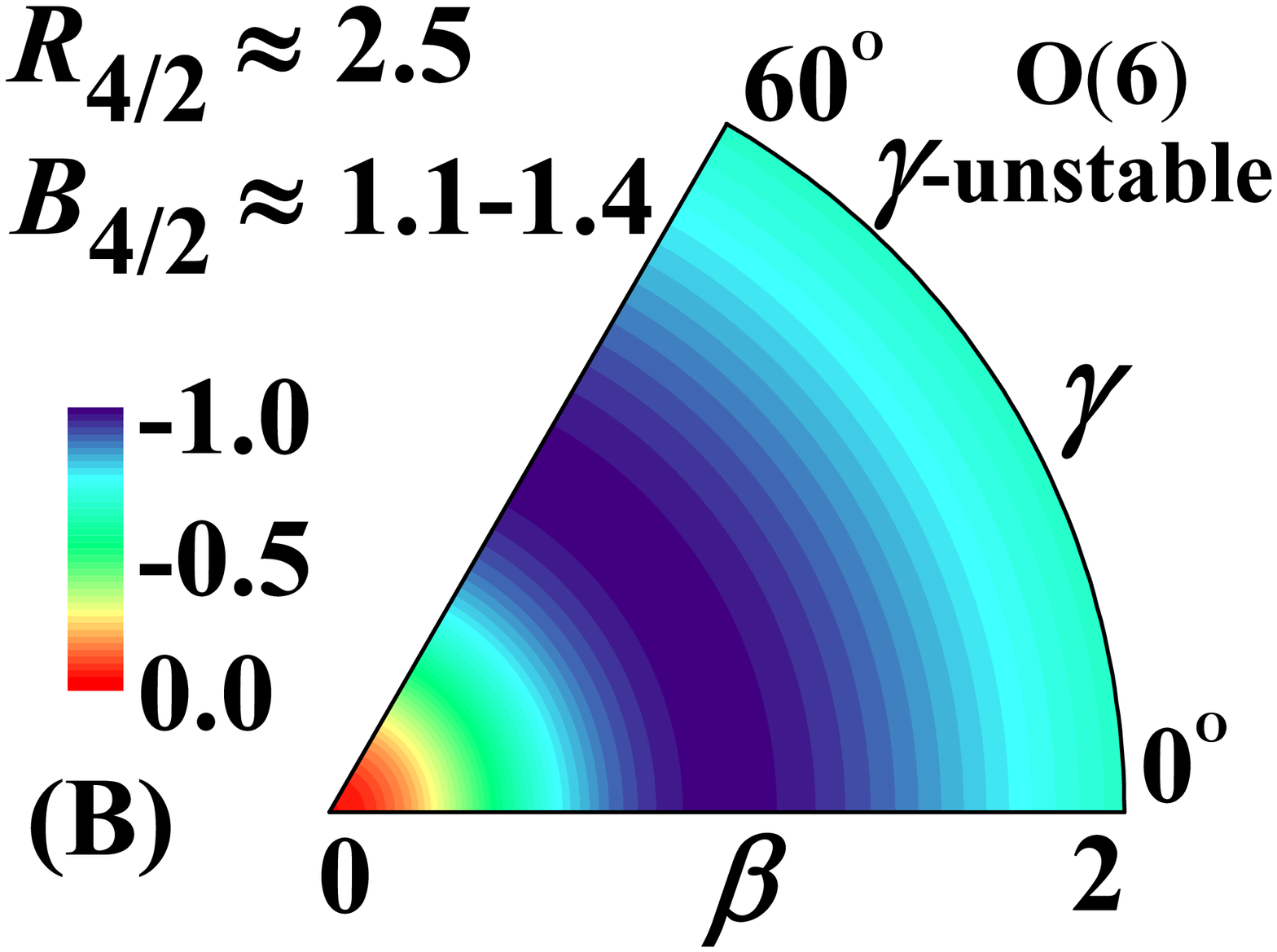}
\includegraphics[scale=0.21]{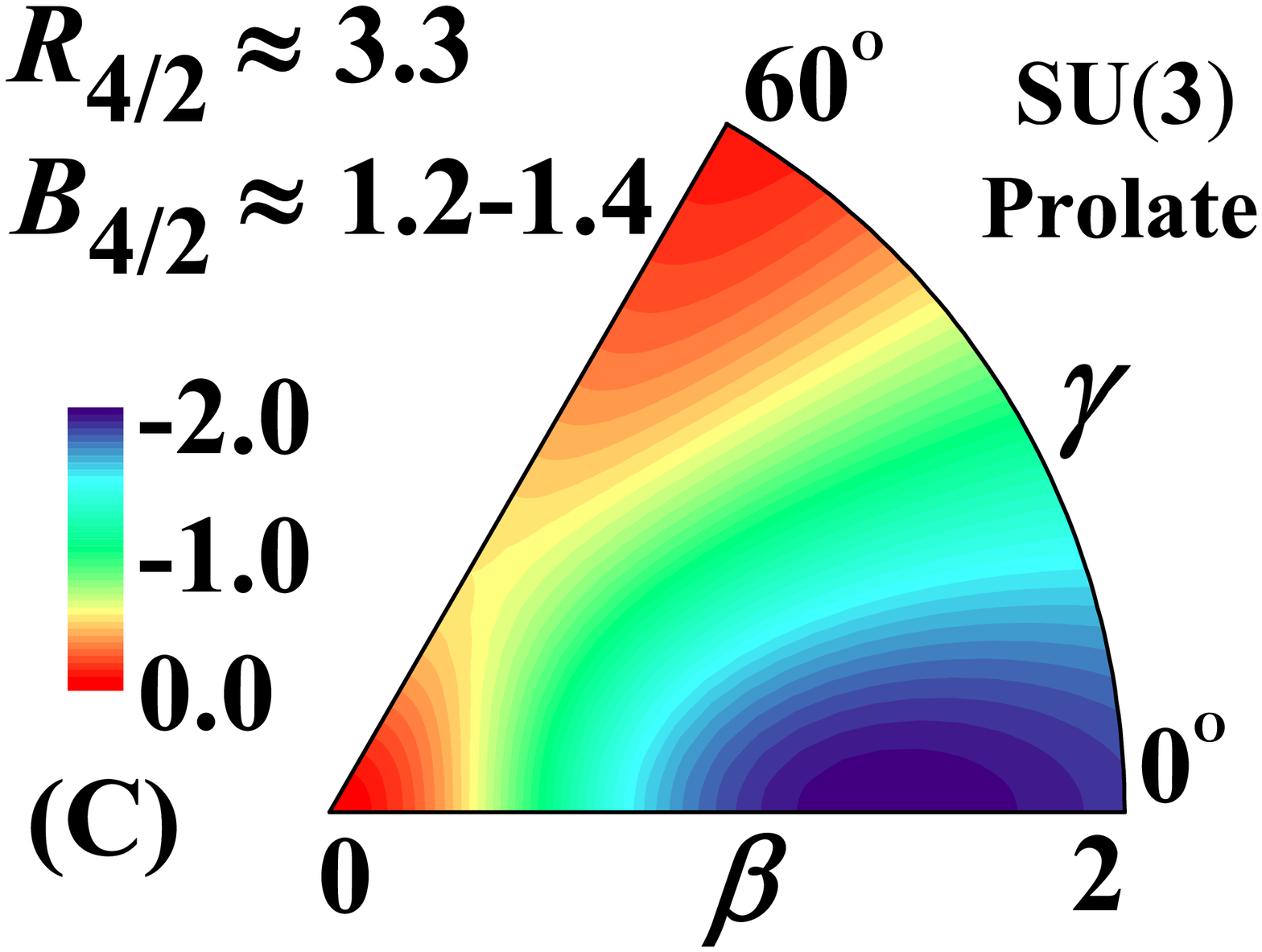}
\includegraphics[scale=0.21]{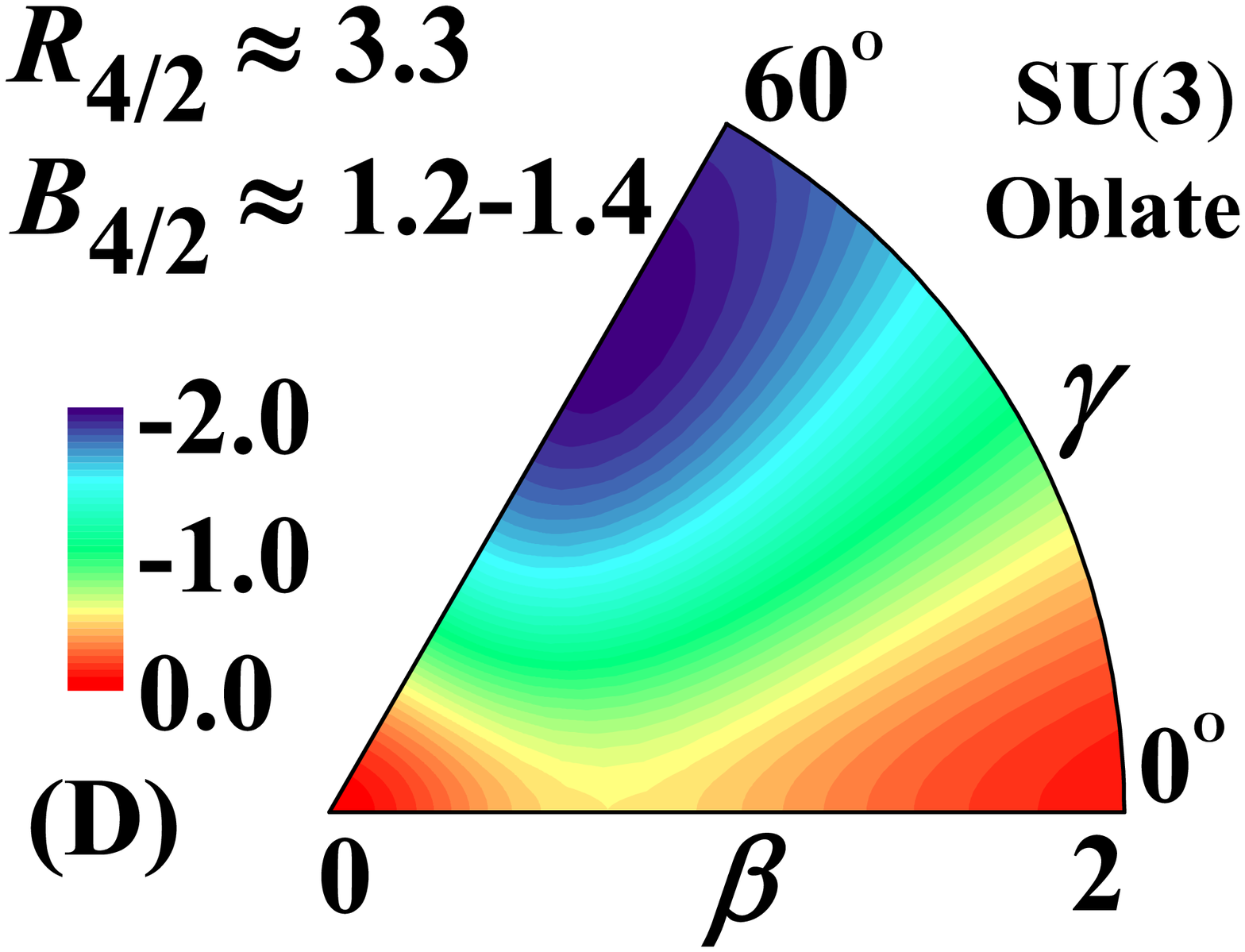}
\caption{Potentials \& Ratios ($R_{4/2}$ \& $B_{4/2}$) produced from (\ref{CQ}) with nonzero parameters
taken as $\varepsilon=1.0$ for U(5), $\kappa=-1.0$ for O(6),
$\kappa=-1.0$ and $\chi=\pm\sqrt{7}/2$ for SU(3) (prolate \& oblate shapes).}\label{F1}
\end{center}
\end{figure}
\vskip .3cm

To describe the conventional
collective modes within the IBM, we adopt the well-known
consistent-$Q$ Hamiltonian~\cite{Warner1983}
\begin{eqnarray}\label{CQ}
\hat{H}_{\mathrm{CQ}}=&\varepsilon\,\hat{n}_d+\kappa\frac{1}{N}\hat{Q}^\chi\cdot\hat{Q}^\chi
\,
\end{eqnarray}
with $\hat{n}_d=d^\dag\cdot\tilde{d}$ and $\hat{Q}_u^\chi =
(d^{\dag} s + s^{\dag} \tilde{d})_u^{(2)} + \chi
(d^{\dag}\times\tilde{d})_u^{(2)}$, where $\varepsilon,~\kappa$ and
$\chi$ are real parameters and $N$ is the total boson number.
Different dynamical symmetries (DSs) in the IBM can be characterized
as: the U(5) when $\varepsilon>0$ and $\kappa=0$; the O(6) when
$\varepsilon=0$, $\kappa<0$ and $\chi=0$; and the SU(3) when
$\varepsilon=0$, $\kappa<0$ and $\chi=\pm\sqrt{7}/{2}$. These DSs in
turn describe the corresponding collective modes in an algebraic
way, including the spherical vibrator (U(5)), $\gamma$-unstable
rotor (O(6)), and axially-deformed rotor (SU(3))~\cite{Jolie2001}.
However, the triaxially-deformed rotor, which is also a typical
collective mode in the Bohr-Mottelson model, is out of reach of the
consistent-$Q$ Hamiltonian.

\vskip .3cm The classical
limit of the IBM Hamiltonian can be worked out by using coherent
state of the system defined as~\cite{IachelloBook87}
\begin{eqnarray}
|\beta, \gamma, N\rangle=N_A[s^\dag + \beta \mathrm{cos} \gamma~
d_0^\dag\ + \frac{1}{\sqrt{2}} \beta \mathrm{sin} \gamma (d_2^\dag +
d_{ - 2}^\dag)]^N |0\rangle\,
\end{eqnarray}
with $N_A=1/\sqrt{N!(1+\beta^2)^N}$.~The classical potential
corresponding to $\hat{H}_{\mathrm{CQ}}$ is then given as
$V(\beta,\gamma)=\frac{1}{N}\langle\beta, \gamma,
N|\hat{H}_{\mathrm{CQ}}|\beta, \gamma,
N\rangle|_{N\rightarrow\infty}$. The potential configurations and the corresponding
ratios of $\hat{H}_{\mathrm{CQ}}$ are shown in
FIG.~\ref{F1}. As expected, different modes indeed exhibit different types of
potential minimum with different $R_{4/2}$ and $B_{4/2}$
ratios. Specifically, it is given by $R_{4/2}\approx2.0$ in the U(5) mode,
$R_{4/2}\approx2.5$ in the O(6) mode, and  $R_{4/2}\approx3.3$
in the SU(3) modes for both prolate and oblate but with different quadrupole moments~\cite{Jolie2001}, of
which the common feature is $R_{4/2}\geq2.0$ and $B_{4/2}>1.0$. It
should also be mentioned that the pairing dominant situation in the shell
model, which belongs to a non-collective mode, may result in
$B_{4/2}\ll1.0$ but accompanied with $R_{4/2}<2.0$. Nevertheless, no triaxial minimum with
$0^\circ<\gamma_\mathrm{min}<60^\circ$ appears in FIG. \ref{F1}
described by the consistent-$Q$ Hamiltonian.

\vskip .3cm To obtain a potential configuration with triaxial
minimum at a mean-field level, higher-order terms have to be introduced
in the IBM~\cite{VC1981}. For example, the stable triaxial minimum
at $\gamma=30^\circ$ can be induced by the cubic term $(d^\dag\times d^\dag\times
d^\dag)^3\cdot(\tilde{d}\times\tilde{d}\times\tilde{d})^3$~\cite{VC1981,Heyde1984}.
A region of triaxiality with $0^\circ<\gamma<60^\circ$ may be allowed in
an extension of the consistent-$Q$ Hamiltonian by adding the cubic
term $(\hat{Q}^\chi\times\hat{Q}^\chi\times\hat{Q}^\chi)^0$~\cite{Fortunato2011}.
However, the existence of a triaxial minimum is insufficient in
producing a triaxially-deformed rotor mode in the IBM. In turn, a
group-guided approach first established in the shell model description of a quantum rotor
~\cite{Leschber1987,Castanos1988} was employed to realize
the triaxial rotor mode in the IBM ~\cite{Smirnov2000}, wherein the
Hamiltonian was constructed with the symmetry-conserving operators
of the $\mathrm{SU(3)}\supset\mathrm{SO(3)}$ integrity
basis~\cite{Vanden1985}. From a group (algebra) theory point of
view~\cite{Ui1970}, the su(3) algebraic relations in the large-$N$
limit contract to those of the semi-simple Lie algebra
$\mathrm{t_5}\oplus\mathrm{so(3)}$ of a quantum rotor, within which an
exact mapping~\cite{Zhang2014} between the triaxial rotor and its
IBM image was established for any $\gamma$-deformation based on the
formulism developed in \cite{Leschber1987,Castanos1988}.

\vskip .3cm In the following, we revisit and reformulate a generic SU(3)-based theory for realizing
a triaxial rotor geometry, one that can be exercised within any application that uses SU(3) basis states, such as the shell model
\cite{Elliott1958,Draayer1983, Draayer2012} and the
IBM~\cite{Smirnov2000,Zhang2014}. In the SU(3) algebraic realization
of a triaxial rotor, the Hamiltonian is divided into its
static and dynamic parts as
\begin{eqnarray}\label{Tri}
\hat{H}_{\mathrm{Tri}}=\hat{H}_\mathrm{S}+\hat{H}_\mathrm{D}\,
\end{eqnarray}
where
\begin{eqnarray}\label{HS}
&\hat{H}_\mathrm{S}=\frac{a_1}{N}\hat{C}_2[\mathrm{SU(3)}]+\frac{a_2}{N^3}\hat{C}_2[\mathrm{SU(3)}]^2+\frac{a_3}{N^2}\hat{C}_3[\mathrm{SU(3)}],\\
\label{HD} &\hat{H}_\mathrm{D}=
t_1\hat{L}^2+t_2(\hat{L}\times\hat{Q}\times\hat{L})^{(0)}+t_3(\hat{L}\times\hat{Q})^{(1)}\cdot(\hat{L}\times\hat{Q})^{(1)}.\
\end{eqnarray}
Here, $\hat{L}$ and $\hat{Q}$ are the angular momentum and
quadrupole momentum operators, respectively, while $a_i$ and $t_i$
with $(i=1,2,3)$ are real parameters.
The SU(3) Casimir operators are defined
as
\begin{eqnarray}\label{C2}
&\hat{C}_2[\mathrm{SU(3)}]=2\hat{Q}\cdot\hat{Q}+\frac{3}{4}\hat{L}^2,\\
\label{C3}
&\hat{C}_3[\mathrm{SU(3)}]=-\frac{4\sqrt{35}}{9}(\hat{Q}\times\hat{Q}\times\hat{Q})_0^{(0)}
-\frac{\sqrt{15}}{2}(\hat{L}\times\hat{Q}\times\hat{L})_0^{(0)}.~~
\end{eqnarray}
It is important to note that by including scalar polynomial forms in $\hat{Q}$ up to $(\hat{Q}\cdot\hat{Q})^2$ the Hamiltonian (\ref{HS}) is able to generate a collective potential of a stable axially-asymmetric system. This feature is consistent with the analysis discussed in \cite{Rowe1985}, except that in the present case the ${Q}$ that is used (\ref{C2}) and (\ref{C3}) are generators of SU(3).
And further, note that scalar polynomial forms that include $\hat{L}$ contribute nothing to the ground state itself, with their action separately or in conjunction with ${Q}$, serving to define the effective moments of inertia of the system. And moreover, eigenvalues of the Casimir operators of SU(3) can be expressed in terms of the SU(3) irreducible representation (irreps) labels $(\lambda,\mu)$; that is,
\begin{eqnarray}
&\langle\hat{C}_2[\mathrm{SU(3)}]\rangle=\lambda^2+\mu^2+3\lambda+3\mu+\lambda\mu,\\
&\langle\hat{C}_3[\mathrm{SU(3)}]\rangle=\frac{1}{9}(\lambda-\mu)(2\lambda+\mu+3)(\lambda+2\mu+3)\,
.
\end{eqnarray}
The static triaxiality
is determined by
$\langle\hat{H}_\mathrm{S}\rangle=f(\lambda,\mu)$
with the $\gamma$-deformation~\cite{Castanos1988}
\begin{eqnarray}\label{gamma}
\gamma_{\,{\mathrm{S}}}=\mathrm{tan}^{-1}(\frac{\sqrt{3}(\mu+1)}{2\lambda+\mu+3})
\,
\end{eqnarray}
and the corresponding $\beta$-deformation
\begin{eqnarray}
\label{beta}
k_0\,\beta_{\,\mathrm{S}}=\sqrt{\lambda^2+\mu^2+\lambda\mu+3\lambda+3\mu+3}
\, ,
\end{eqnarray}
where $k_0$ is a scale factor. The
ground-state energy of
the triaxial Hamiltonian (\ref{Tri}) is
given as $E_g=f(\lambda,\mu)$ at the optimal values
$(\lambda_0,\mu_0)$, with which
the parameters $a_i$ are thus determined,
while $\hat{H}_\mathrm{D}$ contributes nothing to
the ground-state energy. The total Hamiltonian
$\hat{H}_\mathrm{Tri}$ is then applied to generate the SU(3) image
of a triaxial rotor with the Hamiltonian
\begin{eqnarray}\label{rotor}
\hat{H}_{\mathrm{Rot}}=A_1\,\hat{I}_ 1^{~2}+A_2\,\hat{I}_2^{~2}+A_3\,\hat{I}_3^{~2},
\end{eqnarray}
for which the parametes  $t_i$ are fixed by the exact mapping
\cite{Leschber1987, Castanos1988}: $t_i\equiv
t_i(\lambda_0,\mu_0,A_1,A_2,A_3)$. In (\ref{rotor}), $\hat{I}_{\mu}$
are the angular momentum operators in the intrinsic frame and the
inertia parameters $A_i$ can be either extracted from the momentum
of inertia formulas~\cite{Zhang2014} or just taken as independent
parameters~\cite{Wood2004}. For convenience, throughout this work,
$A_1:A_2:A_3=3:1:4$ is taken in the analysis, which corresponds to a
very asymmetric situation. It is important to reiterate that this SU(3)
realization of a triaxial rotor, as shown above, is not restricted to a
specific implementation of SU(3), it is a generic property of the operators
that generate SU(3) regardless of its specific implementation.

\begin{figure}
\begin{center}
\includegraphics[scale=0.21]{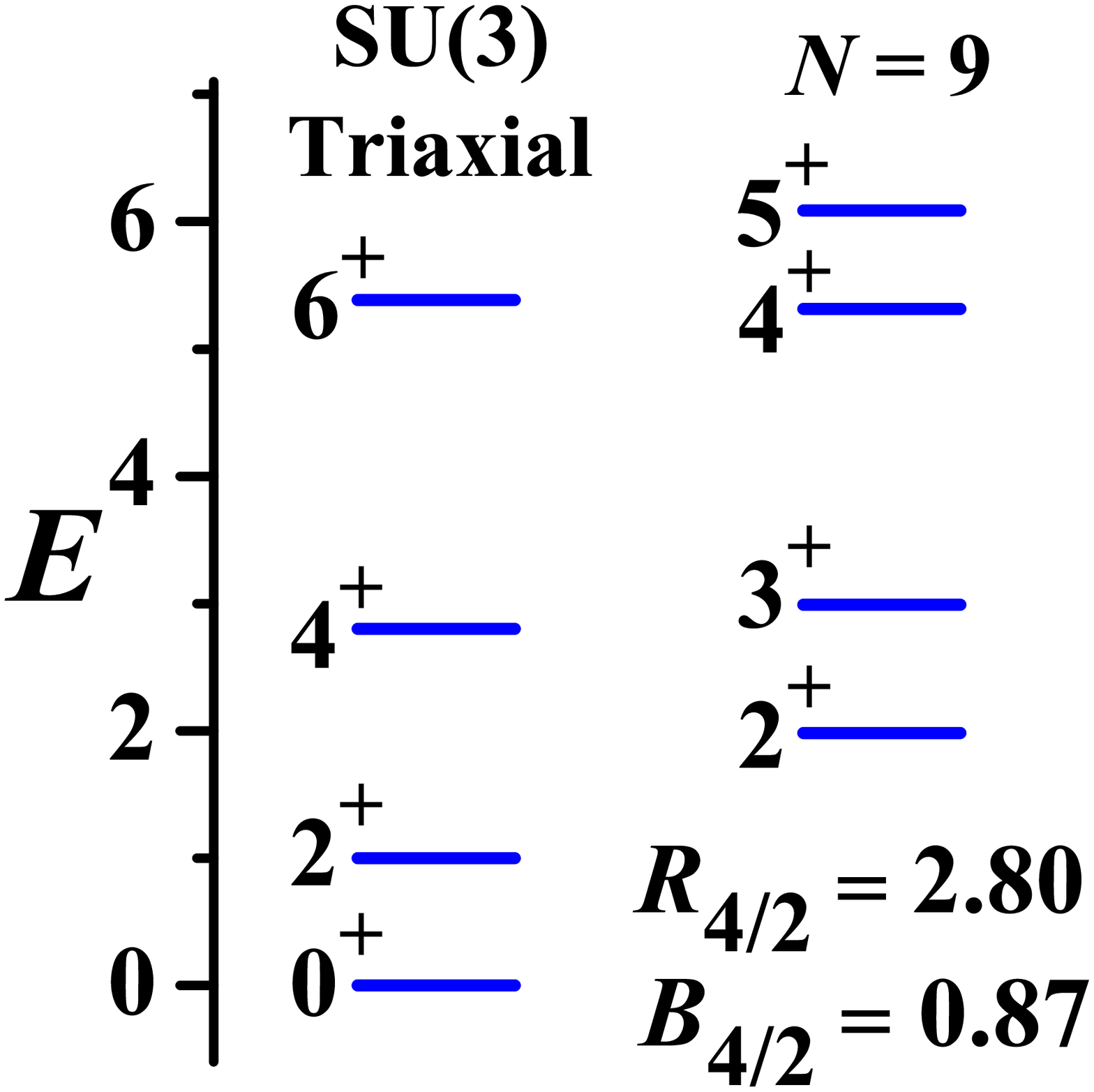}
\includegraphics[scale=0.21]{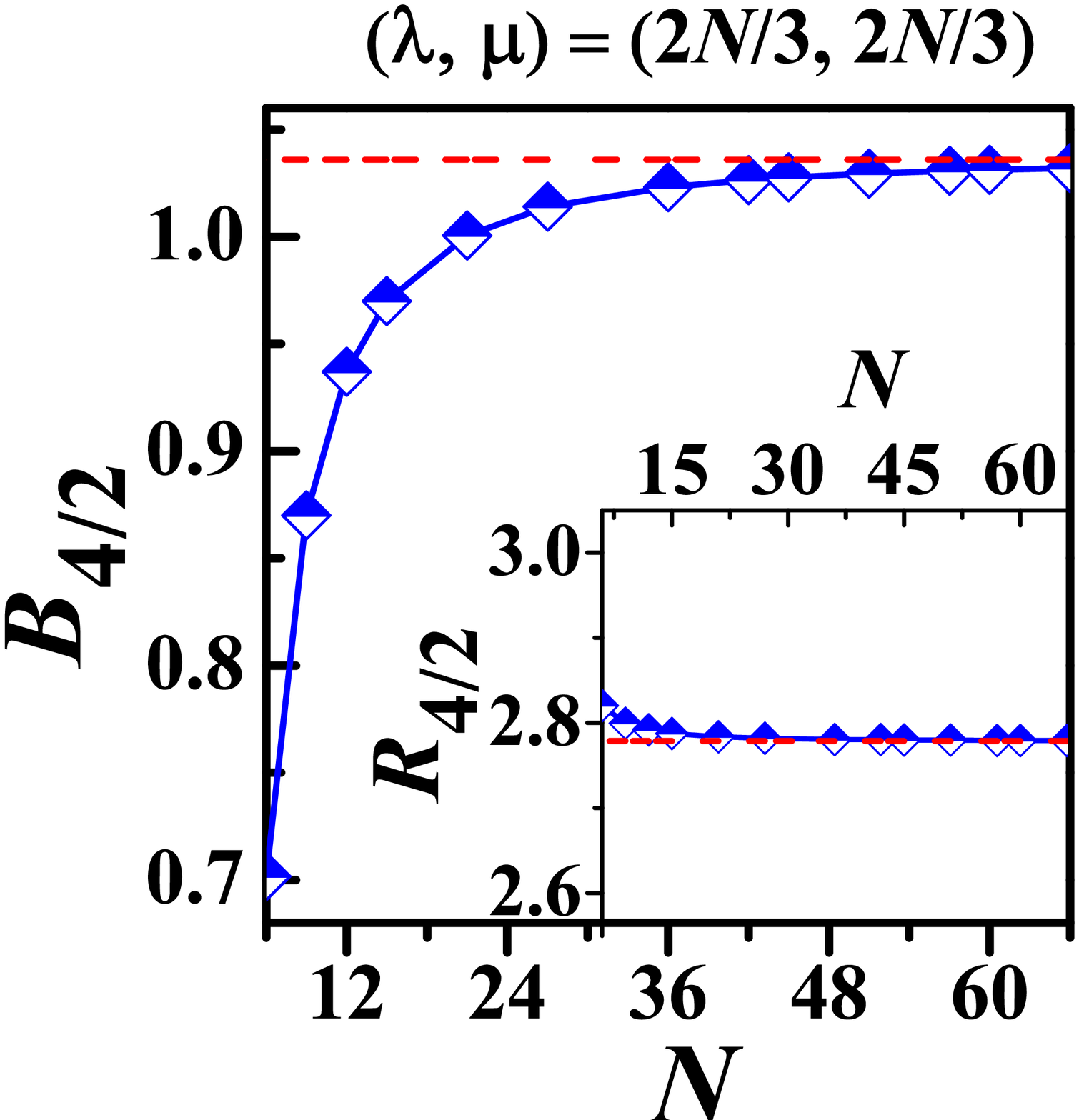}
\caption{The level pattern for $N=9$ with all the level energies normalized to $E(2_1^+)=1.0$
and the evolutions of $B_{4/2}$ and $R_{4/2}$ (inset in the right panel) as
functions of $N$. All the results are obtained from
$\hat{H}_{\mathrm{Tri}}$ with the parameters
$a_1:a_2:a_3=-\frac{27+10N}{3N}:1:1$ generating
$(\lambda_0,\mu_0)=(2N/3,2N/3)$ and
$t_i(\lambda_0,\mu_0,A_1,A_2,A_3)$. The dashed lines in the right panel denote
those obtained directly from the triaxial rotor Hamiltonian
(\ref{rotor}).}\label{F2}
\end{center}
\end{figure}

\vskip .3cm

\begin{figure}
\begin{center}
\includegraphics[scale=0.22]{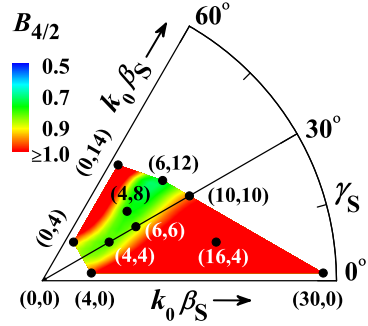}
\caption{Landscape of $B_{4/2}$ with $R_{4/2}\in[2.0,3.33]$ on the $\beta_{~\mathrm{S}}-\gamma_{~\mathrm{S}}$ sector. The
results are obtained from (\ref{HD}) with the parameters mapping from
the triaxial rotor for all given $(\lambda,\mu)$ with
$N\leq15$.}\label{F3}
\end{center}
\end{figure}

In this work, we focus on the IBM realization of a
triaxial rotor. In the IBM, the SU(3) generators are defined as
$\hat{L}_u=\sqrt{10}(d^\dag\times\tilde{d})_u^{(1)}$ and $\hat{Q}_u=\hat{Q}^{\chi}_u$
with $\chi=-\sqrt{7}/2$, which is also taken
as the $E2$ transition operator in the calculation.
We take  $N=9$ as an example to illustrate the finite-$N$ triaxial
rotor mode described by (\ref{Tri}). For this example,
$a_1:a_2:a_3=-\frac{27+10N}{3N}:1:1$ is taken, with which  the SU(3)
irrep of the ground state determined by $\hat{H}_\mathrm{S}$ is
$(\lambda_0,~\mu_0)=(6,~6)$ resulting in
$\gamma_{~{\mathrm{S}}}=30^\circ$ according to (\ref{gamma}). The
other parameters are thus determined by the mapping $t_i\equiv
t_i(\lambda_0,\mu_0,A_1,A_2,A_3)$ with
$t_1=3.0,~t_2=0.553,~t_3=-0.227$ for the $A_1:A_2:A_3$ ratio shown
above. The resulting triaxial structure is shown in the left panel of
FIG.~\ref{F2}, where the low-lying states may group into the standard
rotational bands. The odd-even staggering appearing in the $\gamma$
band confirms the spectrum to be a rigid triaxial rotational one.
Most interestingly, the unusual small $B(E2)$ ratio, $B_{4/2}<1.0$,
appears with the level energies still following the normal
collective excitation value with $R_{4/2}\in[2.0,~3.33]$. Therefore,
it can be concluded that the unusual small $B_{4/2}$ ratio with
normal $R_{4/2}$ ratio occurs in the finite-$N$ triaxial rotor, which applies to other nuclear
models~\cite{Draayer1983,Bonatsos2017,Rowe1985,Georgieva1983,Wu1987,Zuker1995}
with the same triaxial rotor description as well. Moreover, in a
recent work \cite{Wang2020}, small $B_{4/2}$ has been considered to be produced from the $E2$ transition
prohibition between two different irreps in the SU(3) limit of the IBM, in which $0_1^+,~2_1^+$
belong to the SU(3) irrep $(2N, 0)$, while other yrast states with
$L\geq4$ belong to other SU(3) irreps. It is obvious that the
mechanism proposed in this work is completely different from that
of \cite{Wang2020}. To test the finite-$N$ effect in the
present triaxial system, the evolution of $R_{4/2}$ and $B_{4/2}$ as
functions of the boson number $N$ is also worked out and the
results are shown in the right panel of FIG.~\ref{F2}. In the calculation, the
ground-state irrep is chosen to be
$(\lambda_{\,0},~\mu_0)=(2N/3,~2N/3)$ corresponding to
$\gamma_{\,{\mathrm{S}}}=30^\circ$ consistent to the parameter ratio
$a_1:a_2:a_3=-\frac{27+10N}{3N}:1:1$, as the leading SU(3) irrep
with $\lambda+2\mu=2N$ is usually assumed to be dynamically favored.
It can be observed that the $B_{4/2}$ ratio monotonically increases
with the increasing of $N$ from $B_{4/2}<1.0$ to $B_{4/2}>1.0$ and
finally reaches the triaxial rotor  limit value  at very large $N$,
which indicates that the unusual small $B_{4/2}$ ratio in the
triaxial system with $\gamma_{~{\mathrm{S}}}=30^\circ$ occurs mainly
due to the finite-$N$ effect. Meanwhile, the energy ratio $R_{4/2}$
may well coincide with the rotor limit, which confirms the
robustness of the SU(3) mapping procedure. As a more general
asymmetric situation, FIG.~\ref{F3} shows the $B_{4/2}$ evolution on
the $\beta_{~\mathrm{S}}-\gamma_{~\mathrm{S}}$ sector with $N\leq15$, where $k_0$
is taken to be a constant for all SU(3) irreps, in which the
irreps, such as $(2,2)$, with the corresponding $R_{4/2}>3.33$ in the mapping
are excluded. As shown in FIG.~\ref{F3}, though $N$ may be larger than
that used in the previous case, $B_{4/2}<1.0$  survives in the
triaxial region with $\gamma_{~{\mathrm{S}}}\in[25^\circ,45^\circ]$,
which turns to increase toward the rotor limit with further increasing of $N$
consistent with the $N$-dependent behavior shown in the right panel of FIG.~\ref{F2}.

\vskip .3cm

\begin{figure}
\begin{center}
\includegraphics[scale=0.16]{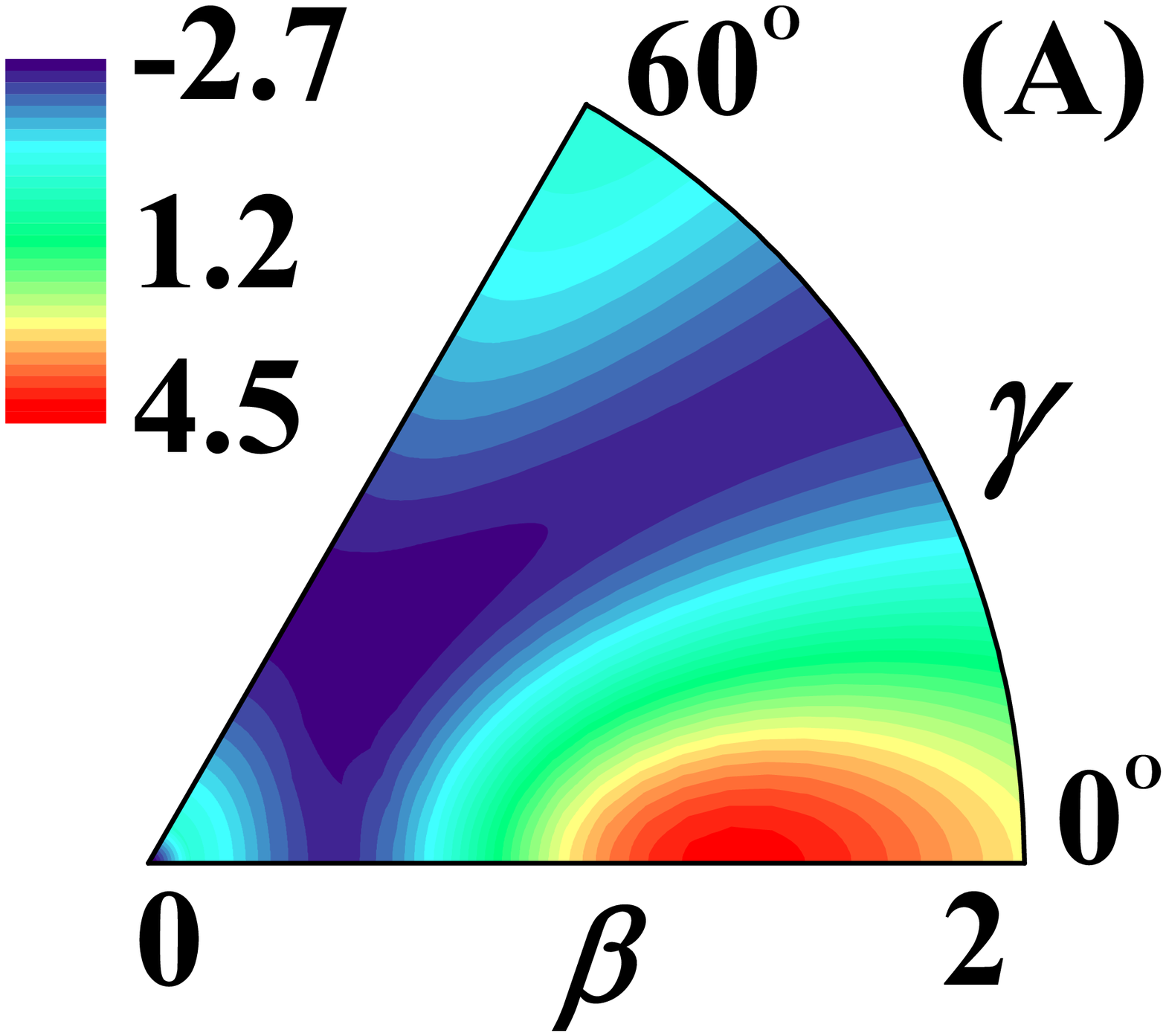}
\includegraphics[scale=0.16]{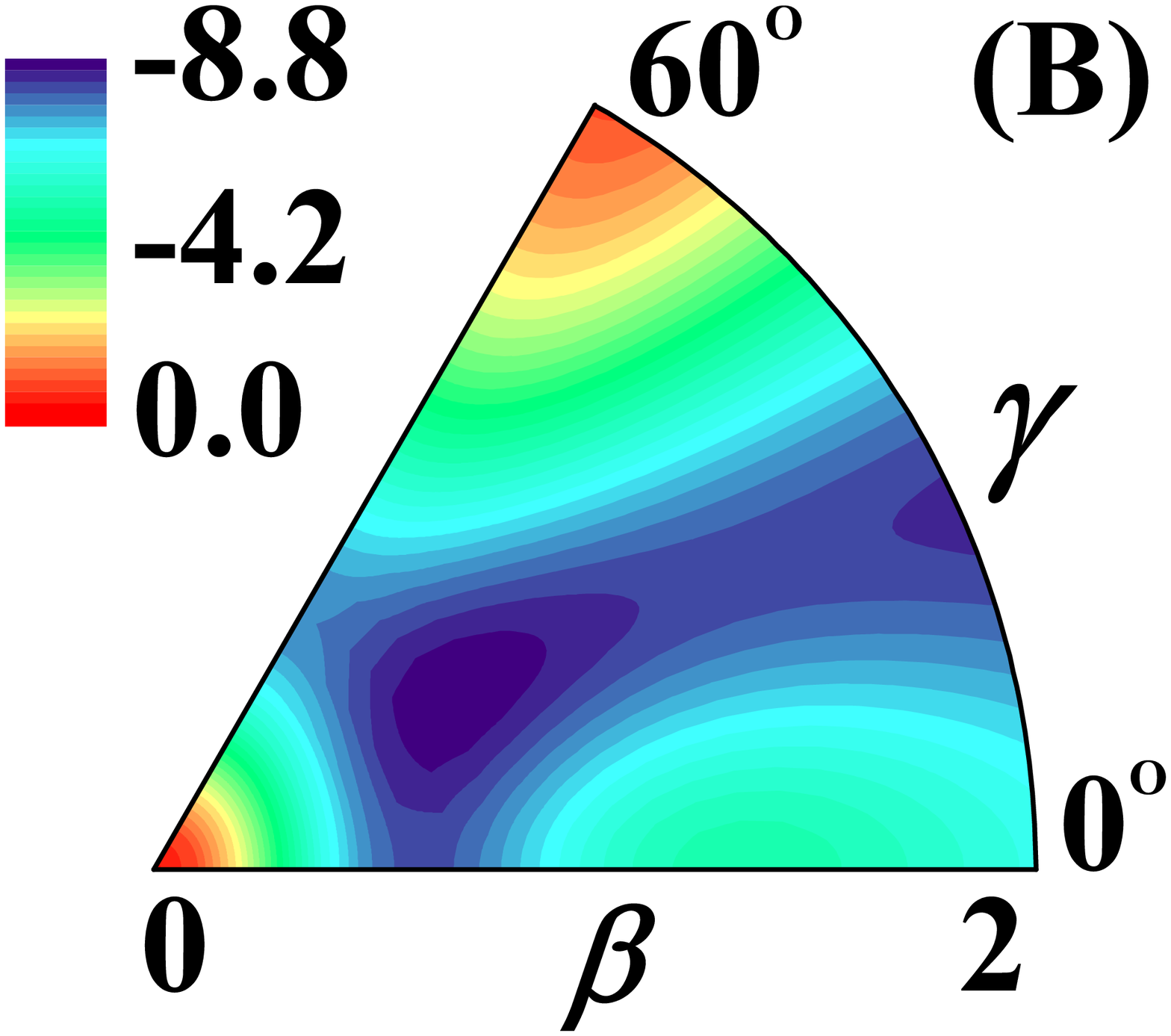}
\includegraphics[scale=0.16]{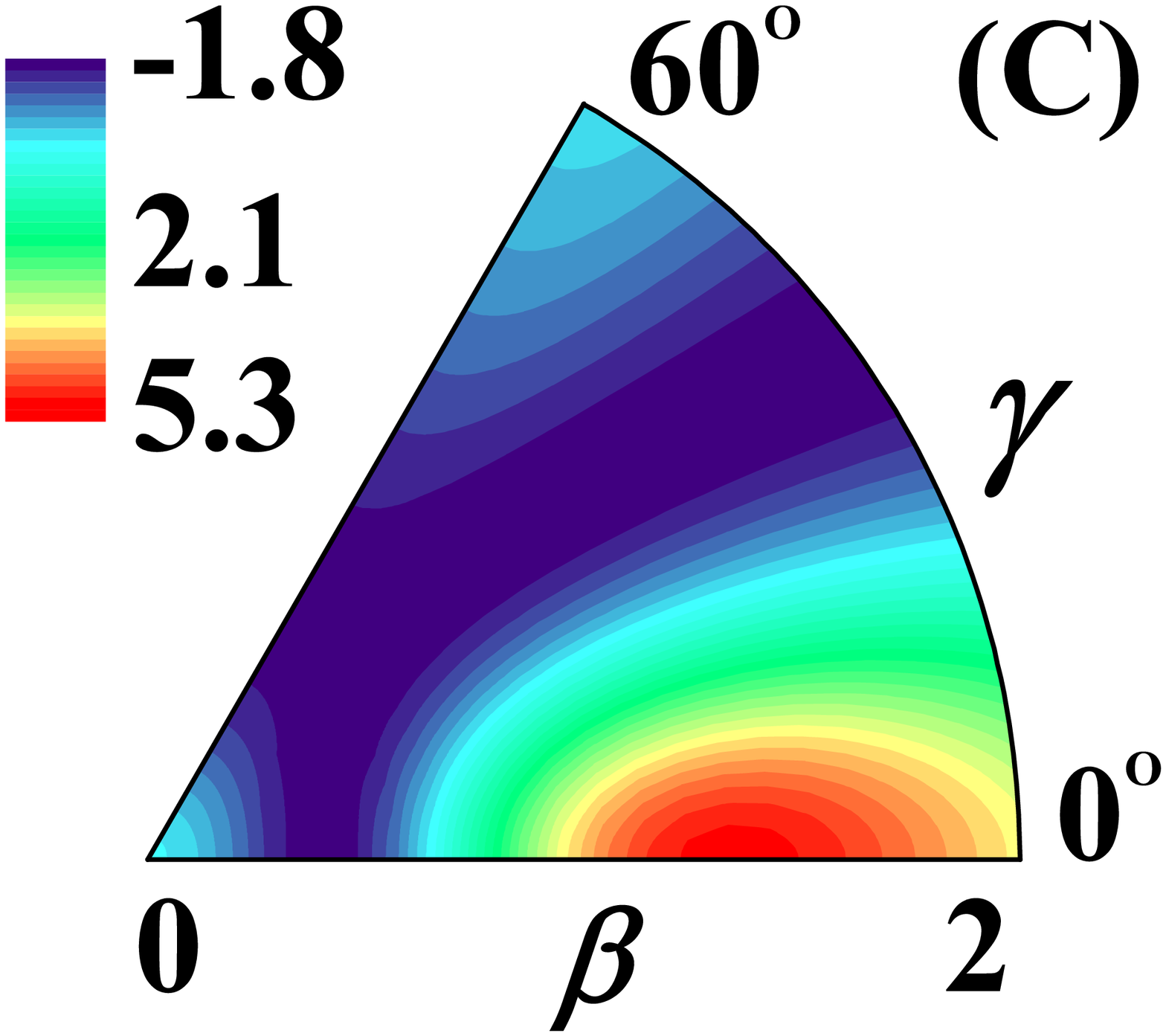}
\includegraphics[scale=0.16]{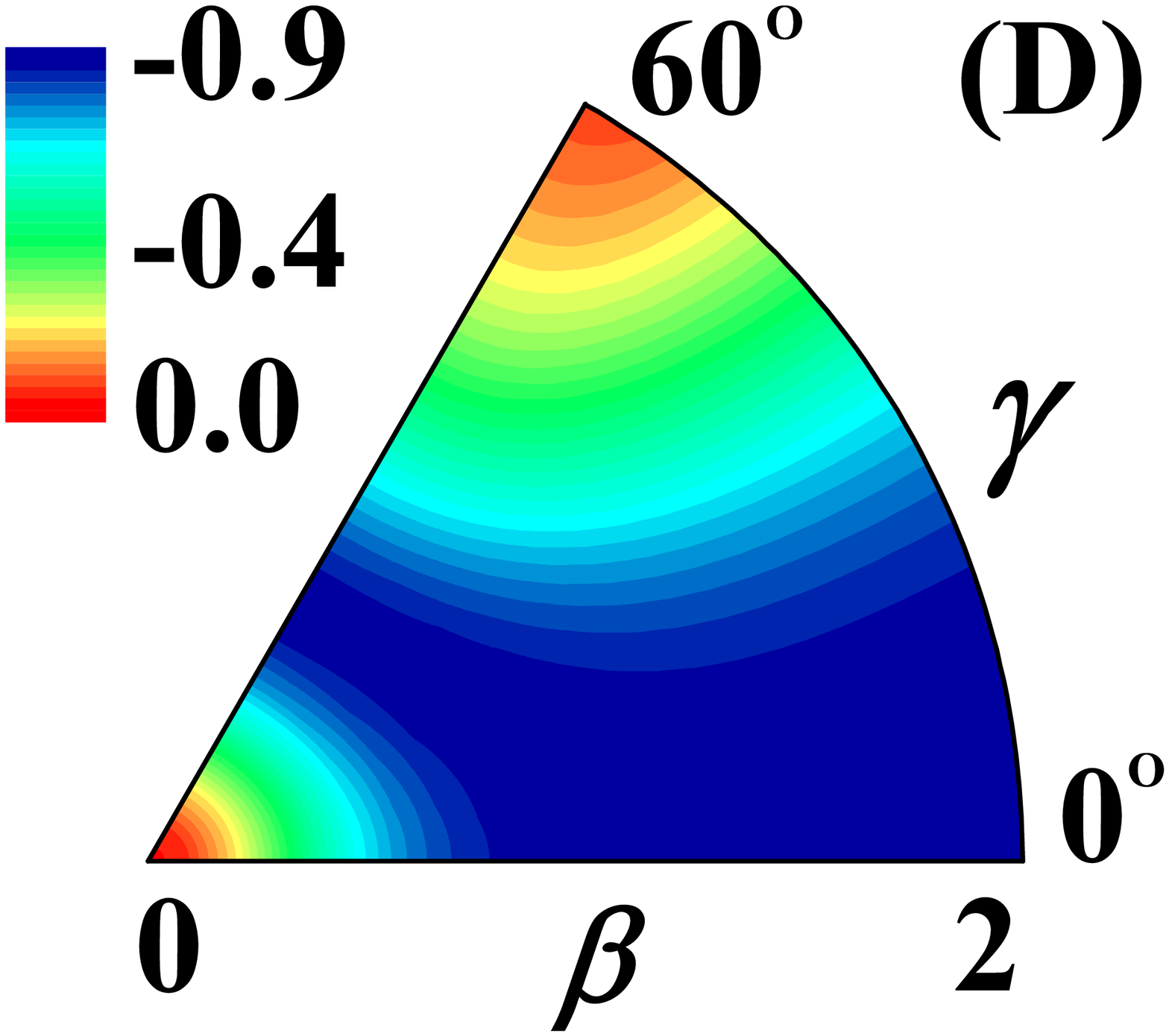}
\caption{Contour plots of (\ref{VS}) with $a_1:a_2:a_3$
given by
(A) $-\frac{27+10N}{3N}:1:1$, (B) $-7.5:1:5$,
(C) $-\frac{24+8N}{3N}:1:0$, and
(D)$-\frac{2N+3}{3N}:0:1$.}\label{F4}
\end{center}
\end{figure}

To reveal the mean-field picture of this triaxial rotor mode,
the potential function for $\hat{H}_\mathrm{S}$
is also calculated by using
the coherent state method~\cite{VC1981}. Specifically,
\begin{eqnarray}\label{VS}\nonumber
V_{\,\mathrm{S}}(\beta,\gamma)&=&\frac{1}{N}\langle\beta, \gamma,
N|\hat{H}_{\mathrm{S}}|\beta, \gamma,N\rangle|_{N\rightarrow\infty}\\
\nonumber
&=&a_1\frac{\beta^2}{(1+\beta^2)^2}\Big[8+4\sqrt{2}\beta\mathrm{cos}(3\gamma)+\beta^2\Big]\\
\nonumber
&+&a_2\frac{\beta^4}{(1+\beta^2)^4}\Big[64+32\beta^2+\beta^4+16\beta^2\mathrm{cos}(6\gamma)\\
\nonumber &+&8\sqrt{2}(8\beta+\beta^3)\mathrm{cos}(3\gamma)\Big]\\
\nonumber
&+&a_3\frac{2\beta^3}{9(1+\beta^2)^3}\Big[24\beta+16\sqrt{2}\mathrm{cos}(3\gamma)\\
 &+&6\sqrt{2}\beta^2\mathrm{cos}(3\gamma)
+\beta^3\mathrm{cos}(6\gamma)\Big]\, .
\end{eqnarray}
The potential $V_{\,\mathrm{S}}(\beta,\gamma)$ simultaneously
describes the classical limit of the triaxial Hamiltonian
$H_{\mathrm{Tri}}$ since its dynamical part
$\langle\hat{H}_\mathrm{D}\rangle/N$ may disappear in the large-$N$
limit through setting an $N$-dependent form of the parameter $t_i$
in (\ref{HD}). One can check that the minimum of
$V_{\,\mathrm{S}}(\beta,\gamma)$ coincides exactly with the
ground-state energy of $\hat{H}_\mathrm{S}/N$ in the large-$N$ limit
and even the $\gamma$-deformation determined by
$V_{\,\mathrm{S}}(\beta,\gamma)$ agrees well with that obtained from
(\ref{gamma}) in the axially-deformed situation. FIG.~\ref{F4} shows
four examples with different $a_1:a_2:a_3$ ratios all generating the
triaxial irrep $(\lambda_0,\mu_0)=(6,6)$ in the $N=9$ case. As shown in
FIG.~\ref{F4}, the potential minimum in panel (A) and (B) with
nonzero $a_{i}$ ($i=1,2,3$) locates at $\gamma\simeq43^\circ$ and
$\gamma\simeq30^\circ$, respectively, while the minima in (C) with
$a_{3}=0$  and (D) with  $a_{2}=0$ are within
$\gamma\in[0^\circ,~45^\circ]$ and $\gamma\in[0^\circ,~60^\circ]$
region, respectively, showing unfixed asymmetric
$\gamma$-deformation in the latter two cases. It means that an
asymmetric deformation can indeed be generated by the static part
$\langle H_{\mathrm{S}}\rangle$, but a stable triaxial minimum can
 be achieved only when both $a_{2}$ and $a_{3}$ are nonzero.
The case (D) actually represents the critical point
situation in the prolate-oblate shape phase transition generated by
a combination of $\hat{C}_2[\mathrm{SU(3)}]$ and
$\hat{C}_3[\mathrm{SU(3)}]$~\cite{Zhang2012}. Clearly, stable
triaxial deformation does not occur even at the critical point
of the prolate-oblate shape phase transition.
Meanwhile, the $\hat{C}_2[\mathrm{SU(3)}]^2$ term is indispensable
in generating a stable triaxial deformation at either finite-$N$
case or large-$N$ limit. Moreover, (\ref{gamma}) gives
$\gamma_{~\mathrm{S}}\simeq30^\circ$ and
$\gamma_{~\mathrm{S}}\simeq14^\circ$ in the large-$N$ limit for case
(A) and (B), respectively. Though the $\gamma_{~\mathrm{S}}$ value
obtained from (\ref{gamma}) is smaller than that obtained from the
coherent state method, the parameter conditions for the triaxial
deformation are consistent with each other. For finite $N$, the
parameter relations shown in (A) and (C) may always guarantee
$(\lambda_0,\mu_0)=(2N/3,2N/3)$ corresponding to
$\gamma_{~\mathrm{S}}=30^\circ$, while those of (D) provides
degenerate ground-state irreps with $\lambda_0+2\mu_0=2N$ for any
$N$ resulting in the prolate-oblate shape (phase) transition due to
level crossing~\cite{Zhang2012}. Furthermore, the finite-$N$
correction to $V_\mathrm{\,S}(\beta,\gamma)$ may result in a very
complicated potential function but only with $1/N$ order
contribution to (\ref{VS}). Therefore,  (\ref{gamma}) seems more
convenient in estimating the $\gamma$-deformation when $N$ is
finite.

\begin{table}
\caption{The model fits for $^{172}$Pt~\cite{Cederwall2018} and
$^{168}$Os~\cite{Grahn2016} with $B(E2;L_i^\pi\rightarrow L_f^\pi)$
values normalized to $B(E2;2_1^+\rightarrow0_1^+)=1.0$, where "-"
indicates the corresponding value is unknown in the experiments.}
\begin{center}
\label{T1}
\begin{tabular}{ccccc|ccc}\hline\hline
$E$(MeV)&$^{172}$Pt&IBM$_a$&IBM$_b$&&$^{168}$Os&IBM$_a$&IBM$_b$\\
\hline
$E(2_1^+)$&0.458&0.458&0.458&&0.341&0.341&0.341\\
$E(4_1^+)$&1.070&1.107&1.070&&0.857&0.872&0.857\\
$E(6_1^+)$&1.753&1.900&1.742&&1.499&1.527&1.504\\
$E(8_1^+)$&2.405&2.549&2.279&&2.222&2.152&2.116\\
$E(2_2^+)$&-&0.749&0.916&&-&0.525&0.686\\
$E(0_2^+)$&-&0.422&0.913&&-&0.287&0.568\\ \hline
Transition&$^{172}$Pt&IBM$_a$&IBM$_b$&&$^{168}$Os&IBM$_a$&IBM$_b$\\
\hline
$2_1^+\rightarrow0_1^+$&1.0&1.0&1.0&&1.0&1.0&1.0\\
$4_1^+\rightarrow2_1^+$&$\mathbf{0.55(19)}$&$\mathbf{0.677}$&$\mathbf{0.552}$&&$\mathbf{0.34(18})$&$\mathbf{0.656}$&$\mathbf{0.351}$\\
$6_1^+\rightarrow4_1^+$&-&0.174&0.131&&-&0.111&0.121\\
$2_2^+\rightarrow0_1^+$&-&0.005&0.201&&-&0.001&0.353\\
$0_2^+\rightarrow2_1^+$&-&0.331&0.007&&-&0.237&0.004\\
\hline\hline
\end{tabular}
\end{center}
\end{table}

\vskip .3cm
To describe realistic nuclear systems,
a more general IBM Hamiltonian with
\begin{equation}\label{H}
\hat{H}=\hat{H}_{\mathrm{CQ}}+\hat{H}_{\mathrm{Tri}}
\end{equation}
may be adopted, which covers all typical collective modes including
spherical vibrator, $\gamma$-unstable, axially-deformed, and
triaxially deformed rotor. Since $\hat{H}_{\mathrm{CQ}}$ includes
SU(3)-symmetry-breaking terms, (\ref{H}) can also be applied to
describe a shape (phase) transitional situation. As a preliminary
application, $^{172}$Pt~\cite{Cederwall2018} and
$^{168}$Os~\cite{Grahn2016}, of which unusual small $B_{4/2}$ ratio
was observed, are fitted by the Hamiltonian (\ref{H}) all with
$N=8$. Due to the scarcity of experimental data, the model
parameters have been fully constrained to the exact mapping from the
triaxial rotor. Here, the mapping is more of a guide for choosing
parameters than just assuming the system as an ideal triaxial rotor.
Specifically, the maximally triaxial irrep for $N=8$ is
$(\lambda_0,\mu_0)=(4,~6)$, which is generated by setting
$a_1:a_2:a_3=-4.5:1:1$. The parameters $t_i$ are then completely
determined from the mapping function
$t_i(\lambda_0,\mu_0,A_1,A_2,A_3)$ with $A_i$ in the triaxial rotor
model being fixed through reproducing the low-lying level energies
of these two nuclei. For simplicity, only the $\hat{n}_d$ term in
$\hat{H}_{\mathrm{CQ}}$ is taken. With these constraints, the
parameters (in MeV) for $^{172}$Pt ($^{168}$Os) are taken as
$\varepsilon=0.049~(0.022),~a_1=-1.21~(-0.768),~a_2=0.27~(0.172),~
a_3=0.27~(0.172),~t_1=0.135~(0.094),~t_2=0.063~( 0.053)$ and
$t_3=0.0015~(0.0026)$, which are denoted by IBM$_a$ in TABLE
\ref{T1}. As shown in TABLE \ref{T1}, the available data for the two
nuclei can be well described by the Hamiltonian (\ref{H}) involving
the triaxial mode with the unusual small $B_{4/2}$ being well
reproduced. Meanwhile, a relatively large $\gamma$-deformation in
the ground state of both nuclei is yielded with
$\langle\gamma_{~\mathrm{S}}\rangle_\mathrm{g}\simeq35^\circ$.
Furthermore, it is shown that the experimental data can be fitted
better if releasing constraints on the model parameters $a_i$ and
$t_i$, of which an example of the fit labelled as IBM$_b$ is also
provided in TABLE \ref{T1} with the model parameters (in MeV)
$\varepsilon=0.186(0.099),~a_1=-0.594(-0.403),~a_2=0.442(0.284),~a_3=0.367(0.226),~t_1=0.020(0.026),~t_2=0.057(
0.043)$ and $t_3=0.0061(0.0056)$ for $^{172}$Pt ($^{168}$Os). Even
though there is rather limited data available to fix the model
parameters, the present analysis clearly shows that triaxial
deformation may occur in these neutron-deficient nuclei, which
agrees to the conclusion made from the mean-field
calculations~\cite{Goasduff2019,Guzman2010}. In addition, it is
worth mentioning that similar collective mode with $B_{4/2}<1.0$
may also appear in intermediate-mass nuclei~\cite{Kintish2014} and
even in light nuclei~\cite{Tobin2014}, for example in $^{20}$Mg and its
mirror partner $^{20}$O. As shown in \cite{Tobin2014}, the SU(3)
irreps involved in the low-lying yrast states of $^{20}$Mg  and
$^{20}$O are mainly dominated by $(\lambda,\mu)=$(4,2) and (6,2) in
the no-core symplectic shell model (NCSpM) description. According to
the present analysis, small triaxial $(\lambda,\mu)$ may result in
$B_{4/2}<1.0$, which agrees not only  to the NCSpM description shown
in~\cite{Tobin2014}, but also to the possible triaxiality in
$^{20}$Mg analyzed in \cite{Mitra2002}.

\vskip .3cm

In summary, it is shown that a collective mode with
$R_{4/2}>2.0$ and $B_{4/2}<1.0$ emerges naturally in the SU(3) realization
of a triaxial rotor in the IBM framework. The mean-field
analysis shows that the $\gamma$-deformation can be induced by the
SU(3) symmetry-conserving terms in the classic
limit~\cite{Heyde1984}, which build the intrinsic configuration for
the algebraic realization of the triaxial rotor. The finite-$N$
effect suppresses the $B_{4/2}$ ratio but keeps the $R_{4/2}$ ratio
nearly unchanged, so that the mode with $B_{4/2}<1.0$ appears when
the low-lying yrast states are dominated by SU(3) irreps with small
$(\lambda,\mu)$. As a preliminary application, the low-lying level
energies and related B($E2$) values of $^{172}$Pt and $^{168}$Os are
calculated from the IBM Hamiltonian involving the triaxial mode. It
is shown that the yrast band and the depressed $B_{4/2}$ ratio can
be excellently reproduced from the model
calculation. The finite-$N$
triaxial rotor mode proposed provides a simple yet promising
mechanism of the anomalous $B(E2)$ values in the yrast band of
neutron-deficient nuclei.
Since the triaxial mode can also be realized in other models similarly
under the SU(3) basis
~\cite{Draayer1983,Bonatsos2017,Rowe1985,Georgieva1983,Wu1987,Zuker1995},
the mechanism proposed in this work is expected to be solid.

\vskip .2cm

\begin{acknowledgments}
Support from the National Natural Science Foundation of China
(11875158, 12175097) and the US National Science Foundation
(PHY-1913728)  is
acknowledged.
\end{acknowledgments}


\bigskip
\begin{thebibliography}{99}

\bibitem{Bohrbook}A. Bohr and B. R. Mottelson, {\it Nuclear Structure II} (Benjamin, New York, 1975).

\bibitem{DF}A. S. Davydov and G. F. Filippov, Nucl. Phys. {\bf 8}, 237 (1958).

\bibitem{WJ}L. Wilets and M. Jean, Phys. Rev. {\bf 102}, 788 (1956).

\bibitem{IachelloBook87}F. Iachello and A. Arima, {\it The Interacting Boson
Model} (England: Cambridge University, 1987).

\bibitem{Elliott1958}J. P. Elliott, Proc. R. Soc. A {\bf245},
128 (1958); {\bf245},
562 (1958); J. P. Elliott and M. Harvey, Proc. R. Soc. A
{\bf272}, 557 (1963); J. P. Elliott and C. E. Wilsdon, Proc. R. Soc.
A {\bf302}, 509 (1968).


\bibitem{Draayer2012}J. P. Draayer, T. Dytrych, K. D. Launey, and D. Langr, Prog. Part. Nucl. Phys. {\bf 67}, 516 (2012).

\bibitem{Barrett2013}B. R. Barrett, P. Navr{\'a}til, and J. P. Vary, Prog. Part. Nucl. Phys. {\bf 69}, 131 (2013).

\bibitem{Dytrych2013}T. Dytrych, K. D. Launey, J. P. Draayer, P. Maris, J. P. Vary, E. Saule, U. Catalyurek, M. Sosonkina,
D. Langr, and M. A. Caprio, Phys. Rev. Lett. {\bf 111}, 252501
(2013).

\bibitem{Dytrych2020}T. Dytrych, K. D. Launey, J. P. Draayer, D. J. Rowe, J. L. Wood, G. Rosensteel, C.
Bahri, D. Langr, and R. B. Baker, Phys. Rev. Lett. {\bf 124}, 042501
(2020).

\bibitem{McCoy2020}A. E. McCoy, M. A. Caprio, T. Dytrych, and P. J. Fasano, Phys. Rev. Lett. {\bf 125},
102505 (2020).

\bibitem{Draayer1983}J. P. Draayer and K. J. Weeks, Phys. Rev. Lett. {\bf51},
1422 (1983); J. P. Draayer, S. C. Park, and O. Casta{\~n}os, Phys.
Rev. Lett. {\bf62}, 20 (1989).

\bibitem{Bonatsos2017}D. Bonatsos, I. E. Assimakis, N. Minkov, A. Martinou, R. B. Cakirli, R. F. Casten, and
K. Blaum, Phys. Rev. C {\bf95}, 064325 (2017); D. Bonatsos, I. E. Assimakis, N.
Minkov, A. Martinou, S. Sarantopoulou, R. B. Cakirli, R. F. Casten,
and K. Blaum, Phys. Rev. C {\bf95}, 064326 (2017).


\bibitem{Grahn2016}T. Grahn {\it et al.}, Phys. Rev. C {\bf 94}, 044327 (2016).

\bibitem{Saygi2017}B. Say{\v g}{\i} {\it et al.}, Phys. Rev. C {\bf 96}, 021301(R) (2017).

\bibitem{Cederwall2018}B. Cederwall {\it et al.}, Phys. Rev. Lett. {\bf 121}, 022502 (2018).


\bibitem{Goasduff2019}A. Goasduff {\it et al.}, Phys. Rev. C {\bf 100}, 034302 (2019).

\bibitem{Zhang2021}W. Zhang {\it et al.}, Phys. Lett. B {\bf 820}, 136527 (2021).

\bibitem{Rowe1985}D. J. Rowe, Rep. Prog. Phys. {\bf48}, 1419 (1985); G. Rosensteel and D. J. Rowe, Ann. Phys. {\bf126},
343 (1980).


\bibitem{Georgieva1983}A. Georgieva, P. Raychev, and R. Roussev, J. Phys. G {\bf9}, 521 (1983).

\bibitem{Wu1987}C. L. Wu, D. H. Feng, X. G. Chen, J. Q. Chen, and M. W.
Guidry, Phys. Rev. C {\bf36}, 1157 (1987).

\bibitem{Zuker1995}A. P. Zuker, J. Retamosa, A. Poves, and E. Caurier, Phys. Rev. C {\bf52}, R1741 (1995).

\bibitem{Warner1983}D. D. Warner and R. F. Casten, Phys. Rev. C {\bf 28}, 1798 (1983).

\bibitem{Jolie2001}J. Jolie, R. F. Casten, P. von Brentano, and V. Werner, Phys. Rev. Lett. {\bf87}, 162501
(2001).

\bibitem{VC1981}P. Van Isacker and J. Q. Chen, Phys. Rev. C {\bf24}, 684 (1981).


\bibitem{Heyde1984}K. Heyde, P. Van Isacker, M. Waroquier and J. Moreau, Phys. Rev. C {\bf29}, 1420 (1984).


\bibitem{Fortunato2011}L. Fortunato, C. E. Alonso, J. M. Arias, J. E. Garc{\'i}a-Ramos, and A.
Vitturi, Phys. Rev. C {\bf 84}, 014326 (2011).

\bibitem{Leschber1987}Y. Leschber and J. P. Draayer, Phys. Lett. B {\bf190}, 1 (1987).

\bibitem{Castanos1988}O. Casta{\~n}os, J. P. Draayer, and Y. Leschber, Z. Phys. A {\bf329}, 33
(1988).

\bibitem{Smirnov2000} Y. F. Smirnov, N. A. Smirnova, and P. Van Isacker, Phys. Rev.
C {\bf61}, 041302(R) (2000).

\bibitem{Vanden1985} G. Vanden Berghe, H. E. De Meyer, and P. Van Isacker, Phys. Rev.
C {\bf32}, 1049 (1985).

\bibitem{Ui1970} H. Ui, Prog. Theor. Phys. {\bf44}, 153 (1970).

\bibitem{Zhang2014}Y. Zhang, F. Pan, L. R. Dai, and J. P. Draayer, Phys. Rev. C {\bf 90}, 044310 (2014).


\bibitem{Wood2004}J. L. Wood, A. M. Oros-Peusquens, R. Zaballa, J. M. Allmond,
and W. D. Kulp, Phys. Rev. C {\bf 70}, 024308 (2004).

\bibitem{Wang2020}T. Wang, EPL {\bf129}, 52001 (2020).

\bibitem{Zhang2012}Y. Zhang, F. Pan, Y. X. Liu, Y. A. Luo, and J. P. Draayer, Phys. Rev. C {\bf 85}, 064312 (2012).

\bibitem{Guzman2010}R. Rodr{\'i}guez-Guzm{\'a}n, P. Sarriguren, L. M. Robledo, and J. E. Garc{\'i}a-Ramos, Phys. Rev. C {\bf 81}, 024310 (2010).

\bibitem{Kintish2014}D. Hertz-Kintish, L. Zamick, and S. J. Q. Robinson, Phys. Rev. C {\bf 90}, 034307 (2014).

\bibitem{Tobin2014}G. K. Tobin, M. C. Ferriss, K. D. Launey, T. Dytrych, J. P. Draayer, A. C. Dreyfuss, and C. Bahri, Phys. Rev. C {\bf 89}, 034312 (2014).

\bibitem{Mitra2002}P. Mitra, G. Gangopadhyay, and B. Malakar, Phys. Rev. C {\bf 65}, 034329 (2002).

\end{thebibliography}
\end{document}